\documentclass[twocolumn,showpacs,preprintnumbers,amsmath,amssymb,aps,prx]{revtex4}

\usepackage{CJK}
\usepackage{array}
\usepackage{txfonts}
\usepackage{amsmath}
\usepackage{graphicx}
\usepackage{dcolumn}
\usepackage{bm}
\usepackage{ulem}

\begin{document}
\begin{CJK*}{GB}{kai}
\CJKfamily{gbsn}

\title{Entanglement of Vector-Polarization States of Photons}

\author{Ling-Jun Kong,$^1$ Yongnan Li,$^1$ Yu Si,$^1$ Rui Liu,$^1$ Zhou-Xiang Wang,$^1$Chenghou Tu,$^1$ and Hui-Tian Wang$^{1,2,}$}

\email{htwang@nankai.edu.cn\htwang@nju.edu.cn}

\affiliation{$^1$School of Physics and Key Laboratory of Weak-Light
Nonlinear Photonics, Nankai University, Tianjin 300071, China \\
$^2$National Laboratory of Solid State Microstructures, Nanjing
University, Nanjing 210093, China}

\date{July 4, 2014}

\begin{abstract}
\noindent
Photons may have homogeneous polarization and may carry quantized orbital angular momentum (OAM). Photon entanglement has been realized in various degrees of freedom such as polarization and OAM. Using a pair of orthogonally polarized states carrying opposite-handedness quantized OAMs could create ``quantized" vector-polarization states with space-variant polarization structures. It is thus possible to extend the polarization degree of freedom from two dimensional space to indefinite dimensional \textit{discrete} Hilbert space. We present a class of vector-polarization entangled Bell states, which use the spatial modes of the vector fields with space-variant polarization structure. We propose a scheme of creating the vector-polarization entangled Bell states using a Sagnac interferometer. We also design an analyzer for identifying the vector-polarization entangled Bell states. Such a class of entanglement is important for quantum information science and technology, and fundamental issues of quantum theory, due to its advantage of the increase in information capacity.
\end{abstract}


\maketitle
\end{CJK*}

%\begin{multicols}

\section{Introduction}

Quantum entanglement, which is one of the quintessential features of quantum theory and a manifestation of nonlocality of quantum mechanics~\cite{R01,R02}, shows stronger correlations than classically explainable~\cite{R03,R04}. It is of great importance for quantum information science and fundamental issues of quantum theory. For example, Bell-basis states as important entangled states have been widely used in quantum information processing tasks such as quantum computation~\cite{R05}, quantum teleportation~\cite{R06,R07,R08}, entanglement swapping~\cite{R09,R10,R11,R12,R13}, quantum dense coding~\cite{R14,R15,R16}, quantum key distribution~\cite{R17,R18}, and quantum secure direct communication~\cite{R19,R20}.

For a two-particle entangled state, neither particle possesses its own well-defined state before performing the measurement, while once the state of one particle is measured the state of another particle is also defined instantaneously. Photon entanglement has be in fact realized in various degrees of freedom. Entangled photon states have hitherto been mostly realized by the two orthogonal polarization states of photons. To enable more efficient use of communication channels in quantum cryptography, multi-dimensional entanglement is of great importance. Higher-order entanglement has been suggested via multiport beam splitters~\cite{R21,R22}.

Orbital angular momentum (OAM) carried by photons with helical phase structures~\cite{R23}, as a fundamentally controllable degree of freedom of photons, has attracted extensive attention and academic interest in quantum foundations and quantum information. The OAM as a degree of freedom has many novel properties and important applications, for instance, uncertainty relations of angular position-OAM~\cite{R24}, violation of the Bell inequality~\cite{R25}, preparation of the four Bell states~\cite{R26}, and qutrit quantum communication protocols~\cite{R27}. As the quanta of OAM carried by a single photon has no theoretical upper limit, which defines an infinitely dimensional discrete Hilbert space, the OAM offers a practical way to create multi-dimensional entanglement~\cite{R28,R29,R30,R31}.

Here we focus on quantum entangled vector-polarization states of photons~\cite{R32,R33}. We present vector-polarization entangled Bell states, which use the spatial modes of the vector fields with space-variant polarization distribution~\cite{R34,R35,R36}. We propose a scheme of creating the vector-polarization entangled Bell states based on a Sagnac interferometer~\cite{R37}. We also design an analyzer for identifying the vector-polarization entangled Bell states. Because the vector fields can be considered as a combination of a pair of orthogonally polarized fields (photons) carrying the opposite-handedness quantized OAMs~\cite{R34}, it is predictable that our approach provides another practical route to define an infinitely dimensional discrete Hilbert space and to future extend to multi-dimensional multi-particle entanglement, enabling more efficient use of communication channels in quantum cryptography. Furthermore, due to the combination of polarization and OAM, the vector-polarization entanglement may increase security with respect to the polarization or OAM entanglement in quantum cryptography. It is conceivable that the vector-polarization Bell states could be of considerable importance in quantum communication, information, cryptography and teleportation, making them versatile and potentially suitable for future technologies.

\section{Vector-Polarization Bell States}

Polarization, momenta and phase are regarded as the controllable degrees of freedom of photons. For the polarization degree of freedom, the maximally entangled Bell states of two photons are written as
\begin{subequations}
\begin{align}\label{01}
| \psi_{\mp}\rangle  = & \frac{1}{\sqrt{2}} \left( | H \rangle | V \rangle \mp | V \rangle | H \rangle \right), \\
| \phi_{\mp}\rangle  = & \frac{1}{\sqrt{2}} \left( | H \rangle | H \rangle \mp | V \rangle | V \rangle \right),
\end{align}
\end{subequations}
\noindent where $| H \rangle$ and $| V \rangle$ indicate the horizontal and vertical polarization eigenstates, respectively.

As is well known, it is easy to transform one Bell state into another one. For instance, for $| \psi_{+}\rangle$, (i) $| \psi_{+}\rangle$ can be changed into $| \phi_{+} \rangle$ with the aid of the polarization exchange ($| H \rangle \Rightarrow | V \rangle $ and  $| V \rangle \Rightarrow | H \rangle $) which can be realized by a half-wave plate (HWP), (ii) $| \psi_{+}\rangle$ can be transformed into $| \psi_{-}\rangle$ with the aid of the polarization-dependent phase shift which is generated by a quarter-wave plate (QWP)~\cite{R15}, and (iii) $| \psi_{+}\rangle$ can be converted into $|\phi_{-}\rangle$ with the aid of both polarization exchange and polarization phase shift, respectively. It should be pointed out that $| V \rangle = \sigma_1 | H \rangle$ and $| H \rangle = \sigma_1 |V \rangle$, where $\sigma_1$ is the well-known Pauli matrix being both Hermitian and unitary, and can also describe the Jones matrix of the HWP with its fast axis at $- \pi/4$ with respect to the horizontal axis.

The optical field with the helical phase structure of $\exp [j (m \varphi + \varphi_0)]$ (shown in the left panel of Fig.~1a) carries the OAM of $ m \hbar$ per photon, where $m$ is referred to as the topological charge or winding number and can take any integer value~\cite{R23,R38}. The OAM degree of freedom can be regarded as a kind of degree of freedom of azimuth-variant phase. Although the helical structure exhibits the phase change in two dimensions in the Cartesian coordinate system, it is in fact the azimuthal phase change in one dimension in the polar coordinate system. During the propagation, the optical fields carrying the OAM exhibit the intertwined (or helical) phase front with its handedness depending on the sign of $m$ (a-1 and a-2 in Fig.~1a). Differently from the SAM or polarization which constructs a two dimensional Hilbert space only, the OAM as a new degree of freedom can define an infinitely dimensional discrete (the quantization of OAM) Hilbert space~\cite{R29,R30,R31}. We have known that the vector fields with the space-variant distribution of states of polarization can be produced by combining a pair of orthogonally polarized optical fields carrying the opposite OAMs (or the helical phases with the opposite handedness)~\cite{R34,R35}. In recent years, the technique of generating the vector fields has become very mature~\cite{R36}. Due to novel properties, the vector fields have many potential applications in various realms~\cite{R39,R40,R41}, including quantum information~\cite{R32,R33}. Since the vector field is always associated with a pair of OAMs with the opposite handedness, it is possible to define an infinitely dimensional discrete (the quantization of vector-polarization state) Hilbert space (Fig.~1b), which provides an opportunity that the vector-polarization states as the degree of freedom are extended from two dimensional Hilbert space to a higher dimensional one. Here we focus on the quantum aspect of the vector-polarization states.

\begin{figure}[bht]
\centerline{\includegraphics[width=8.5cm]{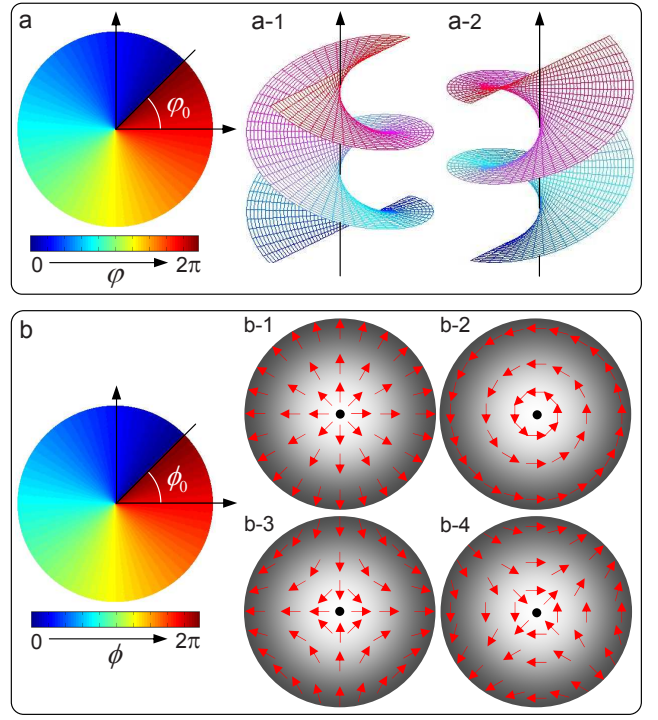}}
\caption{ \textbf{(a)} Vortex optical field with an azimuthal phase term $\exp [j (m \varphi + \varphi_0)]$, carrying an OAM of $m \hbar$, where $\varphi_0$ is the initial phase of an OAM state. The helical phase front (left), and the helical structures of two examples $| m \rangle = |+ 2 \rangle$ (a-1) and $| m \rangle = | -2 \rangle$ (a-2) during the propagation. \textbf{(b)} Polarization distributions of vector-polarization states described by Eqs.~(3) or (4). The helical phase front where $\phi_0$ is the initial phase of a vector-polarization state (left). b-1, b-2, b-3 and b-4 indicate polarization distributions of four vector-polarization states as examples, $| U ^{+1}_R \rangle$, $| U ^{+1}_A \rangle$, $| U ^{-1}_R \rangle$ and $| U ^{-1}_A \rangle$, respectively. $| U ^{+1}_R \rangle$ (b-1) and $| U ^{+1}_A \rangle$ (b-2) for $m = +1$ and $\phi_0 = 0$ are a pair of orthogonal vector-polarization states. $| U ^{-1}_R \rangle$ (b-3) and $| U ^{-1}_A \rangle$ (b-4) for $m = -1$ and $\phi_0 = 0$ are another pair of orthogonal vector-polarization states.}
\end{figure}

We first will construct the vector-polarization entangled Bell states. We then also propose schemes of generating and analyzing the vector-polarization Bell states. For local linearly-polarized vector-polarization states, there are two typical groups of basic states as follows
\begin{subequations}
\begin{align}\label{02}
| U ^{+ m}_R \rangle & = \sin (+ m \phi + \phi_0) | V \rangle + \cos (+ m \phi + \phi_0) | H \rangle, \\
| U ^{+ m}_A \rangle & = \cos (+ m \phi + \phi_0) | V \rangle - \sin (+ m \phi + \phi_0) | H \rangle, \\
| U ^{- m}_R \rangle & = \sin (- m \phi + \phi_0) | V \rangle + \cos (- m \phi + \phi_0) | H \rangle, \\
| U ^{- m}_A \rangle & = \cos (- m \phi + \phi_0) | V \rangle - \sin (- m \phi + \phi_0) | H \rangle,
\end{align}
\end{subequations}
where $m$ is still the topological charge which is the same as in the OAM. We can affirm from Eq.~(2) that ($|U ^{+m}_R \rangle$ and $|U ^{+m}_A \rangle$) and ($|U ^{-m}_R \rangle$ and $|U ^{-m}_A \rangle$) are two pairs of orthogonal vector-polarization states. As examples, when the topological charge $m=1$, four basic vector-polarization states are shown by b-1, b-2, b-3 and b-4 in Fig.~1b, respectively. They are easily interchanged by using two HWPs~\cite{R33}. In fact, the four basic vector-polarization states in Eq.~(2) can be equivalently rewritten as
\begin{subequations}
\begin{align}\label{03}
| U ^{+m}_R \rangle & = \frac{1}{\sqrt{2}} (| R \rangle | - m \rangle + | L \rangle | + m \rangle), \\
| U ^{+m}_A \rangle & = \frac{1}{j \sqrt{2}} (| R \rangle | - m \rangle - | L \rangle | + m \rangle), \\
| U ^{- m}_R \rangle & = \frac{1}{\sqrt{2}} (| R \rangle | + m \rangle + | L \rangle | - m \rangle), \\
| U ^{- m}_A \rangle & = \frac{1}{j \sqrt{2}} (| R \rangle | + m \rangle - | L \rangle | - m \rangle),
\end{align}
\end{subequations}
where $|+ m \rangle$ and $|- m \rangle$ stand for the paraxial spatial modes carrying OAMs of $+ m \hbar$ and $- m \hbar$, and $|R\rangle = \frac{1}{\sqrt{2}} (| H \rangle + j | V \rangle)$ and $| L \rangle = \frac{1}{\sqrt{2}} (| H \rangle - j | V \rangle)$ are the right and left circularly polarized states, respectively. As the forms in Eq.~(3), $| U ^{+m}_R \rangle$, $| U ^{+m}_A \rangle$, $| U ^{- m}_R \rangle$ and $| U ^{- m}_A \rangle$ are also regarded as single-photon ``hybrid" entangled states~\cite{R32}, and any one as a whole state represents a kind of vector-polarization states~\cite{R32}.

Like the traditional polarization Bell states and the OAM Bell states, two-photon vector-polarization Bell states constructed by the orthogonal bases $| U ^{+ m}_R \rangle$ and $| U ^{+ m}_A \rangle$ can be written as follows
\begin{subequations}
\begin{align}\label{04}
| \Psi^{+m}_{-}\rangle  = & \frac{1}{\sqrt{2}} \left( | U ^{+m}_R \rangle | U ^{+m}_A \rangle - | U ^{+m}_A \rangle | U ^{+m}_R \rangle \right), \\
| \Psi^{+m}_{+}\rangle  = & \frac{1}{\sqrt{2}} \left( | U ^{+m}_R \rangle | U ^{+m}_A \rangle + | U ^{+m}_A \rangle | U ^{+m}_R \rangle \right), \\
| \Phi^{+m}_{-}\rangle  = & \frac{1}{\sqrt{2}} \left( | U ^{+m}_R \rangle | U ^{+m}_R \rangle - | U ^{+m}_A \rangle | U ^{+m}_A \rangle \right), \\
| \Phi^{+m}_{+}\rangle  = & \frac{1}{\sqrt{2}} \left( | U ^{+m}_R \rangle | U ^{+m}_R \rangle + | U ^{+m}_A \rangle | U ^{+m}_A \rangle \right).
\end{align}
\end{subequations}
Of course, another pair of orthogonal bases $| U ^{-m}_R \rangle$ and $| U ^{-m}_A \rangle$ can also construct another two-photon vector-polarization Bell states as follows
\begin{subequations}
\begin{align}\label{05}
| \Psi^{-m}_{-}\rangle  = & \frac{1}{\sqrt{2}} \left( | U ^{-m}_R \rangle | U ^{-m}_A \rangle - | U ^{-m}_A \rangle | U ^{-m}_R \rangle \right), \\
| \Psi^{-m}_{+}\rangle  = & \frac{1}{\sqrt{2}} \left( | U ^{-m}_R \rangle | U ^{-m}_A \rangle + | U ^{-m}_A \rangle | U ^{-m}_R \rangle \right), \\
| \Phi^{-m}_{-}\rangle  = & \frac{1}{\sqrt{2}} \left( | U ^{-m}_R \rangle | U ^{-m}_R \rangle - | U ^{-m}_A \rangle | U ^{-m}_A \rangle \right), \\
| \Phi^{-m}_{+}\rangle  = & \frac{1}{\sqrt{2}} \left( | U ^{-m}_R \rangle | U ^{-m}_R \rangle + | U ^{-m}_A \rangle | U ^{-m}_A \rangle \right).
\end{align}
\end{subequations}

\section{Scheme for Preparing Vector-Polarization Bell States}

The scheme of preparing two-photon vector-polarization Bell states, we presented, is shown in Fig.~2. By using degenerate spontaneous parametric down-conversion via noncollinear type-II phase matching in a nonlinear crystal (NLC), the traditional four polarization-entangled Bell states described by Eq.~(1) can be prepared~\cite{R15} (Fig.~2a). The HWP and the QWP in path $p_1$ are used to implement the interchanges among the four polarization-entangled Bell states described in Eq.~(1). Then the polarization-entangled photons $a_1$ and $a_2$ in path $p_1$ and $p_2$ enter into the respective generation systems of vector-polarization states (Fig.~2b), which are composed of a Sagnac interferometer and some optical elements~\cite{R37}. Correspondingly, the polarization-entangled photons $a_1$ and $a_2$ in path $p_1$ and $p_2$ are converted into the vector-polarization entangled photons $b_1$ and $b_2$, respectively. The procedure of preparing the vector-polarization states will be described below.

Before entering the Sagnac interferometer, photons $a_1$ ($a_2$) in path $p_1$ ($p_2$) pass firstly through a HWP whose fast axes has the angle of $-\pi /8$ with respect to the horizontal direction (Fig.~2b). The operator of this HWP, $\mathcal{J}_{HWP}$, can be written by the Jones calculus as
\begin{equation}\label{06}
\mathcal{J}_{HWP} = \frac{1}{\sqrt{2}}
\left[ \begin{array}{cc}
1 & 1\\
1 & -1\end{array} \right].
\end{equation}
This is in fact a Walsh-Hadamard matrix being both Hermitian and unitary. Then the horizontally and vertically polarized components of $a_1$ ($a_2$), separated by the PBS, counter-propagate along a common path in the Sagnac interferometer. A space-variant phase plate (SVPP) with a helical phase front makes the two counter-propagating fields (or photons) feel the opposite chirality of the helical phases of $\exp (\pm j m\phi)$ and then carry the opposite OAMs of $ \pm m \hbar$. Thus the operator of the SVPP can be expressed concisely by the Jones calculus as
\begin{equation}\label{07}
\mathcal{J}_{SVPP} =
\left[ \begin{array}{cc}
e^{+ j m \phi} & 0\\
0 & e^{- j m \phi}\end{array} \right].
\end{equation}
This matrix is always unitary but may not Hermitian except for $m = 0$.

In our scheme (Fig.~2b), a geometric-phase shifter (GPS) composed of three polarization elements (a sandwich structure, i.e. one HWP is sandwiched between two QWPs) is used to control the geometric-phase shift between the two counter-propagating orthogonally polarized components. The fast axes of the two QWPs are parallel each other and are fixed at an angle of $ \pi /4$ with respect to the horizontal direction, while the HWP is allowed to rotate. When the two orthogonally polarized components pass through the GPS, both dynamic phases have no change even though the HWP is rotated~\cite{R42}. Instead, the rotation of the HWP can acquire a controllable geometric-phase shift between the two counter-propagating orthogonally polarized components. The operator of this GPS can be described by the Jones calculus as
\begin{equation}\label{08}
\mathcal{J}_{GPS} (\theta) =
\left[ \begin{array}{cc}
e^{j 2 \theta} & 0\\
0 & - e^{-j 2 \theta}\end{array} \right],
\end{equation}
where $\theta$ is the angle forming the fast axis of the HWP with the horizontal direction. This matrix is always unitary but may not Hermitian except for the cases when $2 \theta$ is an integral multiple of $\pi$. This scheme is robust against the dynamic phase fluctuation caused by the environmental turbulence in such a common closed loop.

After photons leave from the Sagnac interferometer, another QWP whose fast axis forming an angle of $\vartheta$ with the horizontal direction is used to introduce a geometric phase. The operator of the QWP, $\mathcal{J}_{QWP}$, can be described as
\begin{equation}\label{09}
\mathcal{J}_{QWP} (\vartheta) =
\left[ \begin{array}{cc}
\cos ^2 \vartheta - j \sin ^2 \vartheta & - (j + 1) \sin \vartheta \cos \vartheta \\
- (j + 1) \sin \vartheta \cos \vartheta & \sin ^2 \vartheta - j \cos^2 \vartheta \end{array} \right].
\end{equation}
This matrix is always unitary but may not Hermitian except for the cases when $\vartheta$ is an integral multiple of $\pi/2$.

\begin{figure*}[bht]
\centerline{\includegraphics[width=15.0cm]{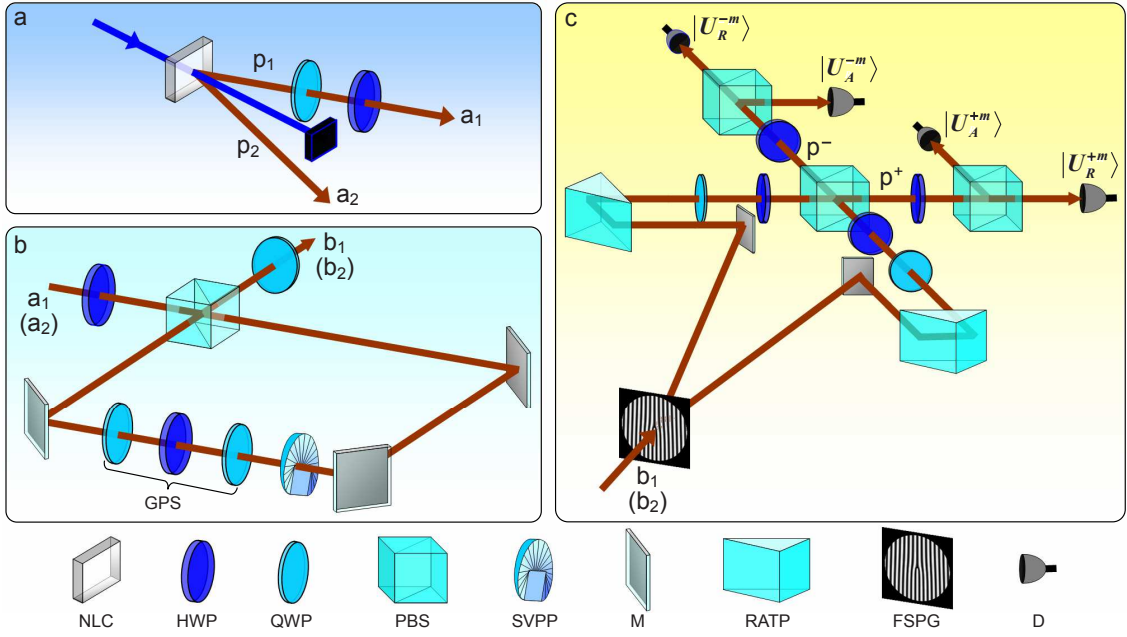}}
\caption{Schemes for preparing and for analysing the vector-polarization Bell states. \textbf{(a)} Scheme for preparing the polarization-entangled states. A continuous wave shorter-wavelength laser can be used to pump a nonlinear crystal (NLC) to produce the polarization-entangled states based on a noncollinear, degenerate type II spontaneous parametric down conversion. By controlling the HWP and QWP, any one of the four polarization-entangled Bell states described by Eq. (1) can be prepared. \textbf{(b)} Scheme for preparing the two-photon vector-polarization Bell states. \textbf{(c)} Scheme for analysing the two-photon vector-polarization Bell states. RATP is a right-angled triangle prism which can be used to adjust the optical length. D is a detect system consisting of single mode fiber and single-photon detector. (See text for details).}
\end{figure*}

With Eqs.~(7) and (8),  the accumulated operator of SVPP and GPS units in the Sagnac interferometers can be rewritten as $\mathcal{J}_{GPS} (\theta) \mathcal{J}_{SVPP}$. After passing through the generation system of vector-polarization states, the evolutions of the eigenstates of $| H \rangle$ and $| V \rangle$ when $\theta = \pi/8$ and $\vartheta = \pi/4$ are as follows
\begin{subequations}
\begin{align}\label{10}
| H \rangle \Rightarrow & \mathcal{J}_{QWP} (\pi/4) \mathcal{J}_{GPS} (\pi/8) \mathcal{J}_{SVPP} \mathcal{J}_{HWP} | H \rangle = e^{+ j \pi /4} | U ^{+m}_R \rangle, \\
| V \rangle \Rightarrow  &\mathcal{J}_{QWP} (\pi/4) \mathcal{J}_{GPS} (\pi/8) \mathcal{J}_{SVPP} \mathcal{J}_{HWP} | V \rangle = e^{- j \pi /4} | U ^{+m}_A \rangle.
\end{align}
\end{subequations}
Then the traditional four polarization Bell states described by Eq.~(1) will be converted into the four vector-polarization Bell states described by Eq.~(4) as $| \psi_{\mp} \rangle \Rightarrow | \Psi^{+m}_{\mp} \rangle$ and $| \phi_{\mp} \rangle \Rightarrow | \Phi^{+m}_{\pm} \rangle$.

When $\theta=\pi/8$ and $\vartheta = -\pi/4$ , the evolutions of the eigenstates $| H \rangle$ and $| V \rangle$ can be written as follows
\begin{subequations}
\begin{align}\label{11}
| H \rangle \Rightarrow  &\mathcal{J}_{QWP} (- \pi/4) \mathcal{J}_{GPS} (\pi/8) \mathcal{J}_{SVPP} \mathcal{J}_{HWP} | H \rangle = e^{+ j \pi /4} | U ^{-m}_A \rangle, \\
| V \rangle \Rightarrow  &\mathcal{J}_{QWP} (- \pi/4) \mathcal{J}_{GPS} (\pi/8) \mathcal{J}_{SVPP} \mathcal{J}_{HWP} | V \rangle = e^{- j \pi/4} | U ^{-m}_R \rangle.
\end{align}
\end{subequations}
In this case, the traditional four polarization Bell states described by Eq.~(1) will be converted into the four vector-polarization Bell states described by Eq.~(5) as $| \psi_{\mp} \rangle \Rightarrow | \Psi^{-m}_{\mp} \rangle$ and $| \phi _{\mp} \rangle \Rightarrow | \Phi^{-m}_{\pm} \rangle$. After leaving from the generation system of vector-polarization states, photons $a_1$ and $a_2$ are converted into photons $b_1$ and $b_2$, respectively. By manipulating the polarization Bell states, therefore, the preparation of the vector-polarization Bell states can be realized.

\section{Analyzer of Vector-Polarization Bell States}

We present a vector-polarization Bell-state analyzer (Fig.~2c) which is similar to the analyzer in Ref.~\cite{R32}. Photons $b_1$ ($b_2$) will meet a fork-shaped phase grating (FSPG). When the FSPG has the same topological charge with the SVPP mentioned above, an incoming photon in the state $ | U_{X}^{+m}\rangle$ $( | U_{X}^{-m}\rangle )$, where $X = R$ or $A$, will be transformed into two equal-probability diffraction beams: One is the right-circularly polarized state $| R \rangle$ in the $+1$st ($-1$st) diffraction order and the other one is the left-circularly polarized state $| L \rangle$ in the $-1$st ($+1$st) diffraction order, and both beams carry no OAM. A QWP and a HWP are then used to transform $| R \rangle$ and $| L \rangle$ into $| V \rangle$ and $| H \rangle$, respectively. The two equal-probability diffraction beams will merge on a PBS. Photons in the states $ | U_{X}^{+m} \rangle$ ($ | U_{X}^{-m} \rangle$) exit in path $p^+$ $(p^-)$ behind the PBS as shown in Fig.~2c. By using two HWPs in paths $p^+$ and $p^-$ (whose fast axes have an angle of $- \pi / 8$ with respect to the horizontal direction), the four vector-polarization states, $| U ^{+m}_R \rangle$, $| U ^{+m}_A \rangle$, $| U ^{- m}_R \rangle$ and $| U ^{- m}_A \rangle$, can be distinguished with other two PBSs. Finally, the analysis of the vector-polarization-entangled Bell states with joint measurement can be realized.

\section{Discussion}

It should be noted that there are some similarities between the OAM states and the vector-polarization states. For example, both of them use the same parameter called topological charge $m$ to describe the two-dimensional spatial distributions, as shown in Fig.~1. However, they are two complete independent degrees of freedom. The OAM states are associated with the phase, while the vector-polarization states are related to the polarization. Photons in the vector-polarization state carry the information of both the OAM and the space-variant polarization. Compared with the OAM states, the vector-polarization states have some advantages such as, for a given topological charge $m$, only one OAM state $| m \rangle$ corresponds to it, while there are two vector-polarization states $| U _R ^{m} \rangle$ and $| U _A ^{m} \rangle$. A major challenge is to ultimately confirm the entanglement in experiment, which may be a Bell inequality experiment generalized to more states. For a linearly-polarized pump field with zero angular momentum (including spin and orbital angular momentum), the emitted state can be represented by
\begin{align}\label{12}
| \psi \rangle = & \sum^{+\infty}_{m=0} (C^{+m,+m}_{R,R} |U^{+m}_R \rangle |U^{+m}_R \rangle + C^{+m,+m}_{R,A} |U^{+m}_R \rangle |U^{+m}_A \rangle \nonumber \\
& \quad + C^{+m,-m}_{R,R} |U^{+m}_R \rangle |U^{-m}_R \rangle + C^{+m,-m}_{R,A} |U^{+m}_R \rangle |U^{-m}_A \rangle \nonumber \\
& \quad + C^{+m,+m}_{A,R} |U^{+m}_A \rangle |U^{+m}_R \rangle + C^{+m,+m}_{A,A} |U^{+m}_A \rangle |U^{+m}_A \rangle \nonumber \\
& \quad + C^{+m,-m}_{A,R} |U^{+m}_A \rangle |U^{-m}_R \rangle + C^{+m,-m}_{A,A} |U^{+m}_A \rangle |U^{-m}_A \rangle \nonumber \\
& \quad + C^{-m,+m}_{R,R} |U^{-m}_R \rangle |U^{+m}_R \rangle + C^{-m,+m}_{R,A}|U^{-m}_R \rangle |U^{+m}_A \rangle \nonumber \\
& \quad + C^{-m,-m}_{R,R} |U^{-m}_R \rangle |U^{-m}_R \rangle + C^{-m,-m}_{R,A} |U^{-m}_R \rangle |U^{-m}_A \rangle \nonumber \\
& \quad + C^{-m,+m}_{A,R} |U^{-m}_A \rangle |U^{+m}_R \rangle + C^{-m,+m}_{A,A} |U^{-m}_A \rangle |U^{+m}_A \rangle \nonumber \\
& \quad + C^{-m,-m}_{A,R} |U^{-m}_A \rangle |U^{-m}_R \rangle + C^{-m,-m}_{A,A} |U^{-m}_A \rangle |U^{-m}_A \rangle ),
\end{align}

\noindent where $C^{i,j}_{P,Q}$ denote the corresponding probability amplitude for measuring $|U^{i}_P \rangle |U^{j}_Q \rangle$ ($i, j = +m, -m$ and $P, Q = R, A$). The photonic state~(12) is an infinite dimensional entangled state for two photons, meaning neither photon possesses a well-defined vector-polarization state after parametric down-conversion. The measurement of one photon defines its vector-polarization state, and projects the second one into the corresponding vector-polarization state. The state in Eq.~(12) is composed of infinite dimensional vector-polarization basis which forms an infinite dimensional Hilbert space.

In summary, we have proposed the concept of the vector-polarization Bell state from traditional polarization state by analogy between the OAM and phase, and extended the Hilbert space of degree of polarization from two dimensions to indefinite dimensions. The polarization entangled Bell states have been constructed. By using some optical elements and a Sagnac interferometer, an effective vector-polarization states generation system has been presented. We have also designed an analyzer to distinguish the vector-polarization Bell states by using the linear optical elements only. The similarities and differences between OAM states and vector-polarization states have been pointed out. The vector-polarization state is an independent degree of freedom from the OAM states although it is always associated to a pair of opposite OAMs. Differently from the traditional polarization states, the vector-polarization states are also ``quantized", because which are always associated with a pair of quantized OAMs with opposite handedness. These vector-polarization states could be in fact extended to higher-dimensional multi-particle entanglement. These vector-polarization states could be anticipated many applications in quantum cryptography, quantum teleportation, quantum communication, and quantum information, due to their higher dimensions and larger flux of information. These photon states should be versatile and potentially suitable for future photonic technologies.

\section*{ACKNOWLEDGMENTS}

This work is supported by the National Basic Research Program (973 Program) of China (No. 2012CB921900), National Natural Science Foundation of China (No. 11274183 and No. 11374166), 111 Project (No. B07013), the National scientific instrument and equipment development project (No. 2012YQ17004), and Tianjin research program of application foundation and advanced technology (No. 13JCZDJC33800 and No. 12JCYBJC10700).


\begin{references}

\bibitem{R01} E. Schr\"{o}dinger, \textit{Die gegenw\"{a}rtige Situation in der Quantenmechanik}, Naturwissenschaften \textbf{23}, 807-812; 823-828; 844-849 (1935).
\bibitem{R02} E. Schr\"{o}dinger, \textit{Discussion of probability relations between separated systems}, Proc. Camb. Phil. Soc. \textbf{31}, 555-563 (1935).

\bibitem{R03} A. Einstein, B. Podolsky, and N. Rosen, \textit{Can quantum-mechanical description of physical reality be considered complete?}, Phys. Rev. \textbf{47}, 777 (1935).

\bibitem{R04} J. S. Bell, \textit{On the Einstein Podolsky Rosen Paradox}, Physics \textbf{1}, 195 (1964).

\bibitem{R05} A. Barenco, D. Deutsch, A. Ekert, and R. Jozsa, \textit{Conditional quantum dynamics and logic gates}, Phys. Rev. Lett. \textbf{74}, 4083 (1995); T. Sleator and H. Weinfurter, \textit{Realizable universal quantum logic gates}, Phys. Rev. Lett. \textbf{74}, 4087 (1995).

\bibitem{R06} C. H. Bennett, G. Brassard, C. Cr\'{e}peau, R. Jozsa, A. Peres, and W. K. Wooters, \textit{Teleporting an unknown quantum state via dual classical and Einstein-Podolsky-Rosen channels}, Phys. Rev. Lett. \textbf{70}, 1895 (1993).

\bibitem{R07} D. Bouwmeester, J. W. Pan, K. Mattle, M. Eibl, H. Weinfurter, and A. Zeilinger, \textit{Experimental quantum teleportation}, Nature (London) \textbf{390}, 575 (1997).

\bibitem{R08} D. Boschi, S. Branca, F. DeMartini, L. Hardy, and S. Popescu, \textit{Experimental realization of teleporting an unknown pure quantum state via dual classical and Einstein-Podolsky-Rosen channels}, Phys. Rev. Lett. \textbf{80}, 1121 (1998).

\bibitem{R09} M. Zukowski, A. Zeilinger, M. A. Horne, and A. K. Ekert, \textit{``Event-ready-detectors" Bell experiment via entanglement swapping}, Phys. Rev. Lett. \textbf{71}, 4287 (1993).

\bibitem{R10} J. W. Pan, D. Bouwmeester, H. Weinfurter, and A. Zeilinger, \textit{Experimental entanglement swapping: entangling photons that never interacted}, Phys. Rev. Lett. \textbf{80}, 3891 (1998).

\bibitem{R11} T. Jennewein, G. Weihs, J. W. Pan, and A. Zeilinger, \textit{Experimental nonlocality proof of quantum teleportation and entanglement swapping}, Phys. Rev. Lett. \textbf{88}, 017903 (2002).

\bibitem{R12} H. de Riedmatten, I. Marcikic, J. A. W. van Houwelingen, W. Tittel, H. Zbinden, and N. Gisin, \textit{Long-distance entanglement swapping with photons from separated sources}, Phys. Rev. A \textbf{71}, 0500302(R) (2005).

\bibitem{R13} A. M. Goebel, C. Wagenknecht, Q. Zhang, Y. A. Chen, K. Chen, J. Schmiedmayer, and J. W. Pan, \textit{Multistage entanglement swapping}, Phys. Rev. Lett. \textbf{101}, 080403 (2008).

\bibitem{R14} C. H. Bennett and S. J. Wiesner, \textit{Communication via one- and two-particle operators on Einstein-Podolsky-Rosen states}, Phys. Rev. Lett. \textbf{69}, 2881 (1992).

\bibitem{R15} K. Mattle, H. Weinfurter, P. G. Kwiat, and A. Zeilinger, \textit{Dense coding in experimental quantum communication}, Phys. Rev. Lett. \textbf{76}, 4656 (1996).

\bibitem{R16} K. Shimizu, N. Imoto, and T. Mukai, \textit{Dense coding in photonic quantum communication with enhanced information capacity}, Phys. Rev. A \textbf{59}, 1092 (1999); X. S. Liu, G. L. Long, D. M. Tong, and F. Li, \textit{General scheme for superdense coding between multiparties}, Phys. Rev. A \textbf{65}, 022304 (2002).

\bibitem{R17} C. H. Bennett, G. Brassard, and N. D. Mermin, \textit{Quantum cryptography without Bell's theorem}, Phys. Rev. Lett. \textbf{68}, 557 (1992); A. K. Ekert, \textit{Quantum cryptography based on Bell's theorem}, Phys. Rev. Lett. \textbf{67}, 661 (1991).

\bibitem{R18} N. Gisin, G. Ribordy, W. Tittel, and H. Zbinden, \textit{Quantum cryptography}, Rev. Mod. Phys. \textbf{74}, 145 (2002).

\bibitem{R19} G. L. Long and X. S. Liu, \textit{Theoretically efficient high-capacity quantum-key-distribution scheme}, Phys. Rev. A \textbf{65}, 032302 (2002).

\bibitem{R20} F. G. Deng and G. L. Long, \textit{Secure direct communication with a quantum one-time pad}, Phys. Rev. A \textbf{69}, 052319 (2004).

\bibitem{R21} M. Reck, A. Zeilinger, H. J. Bernstein, and P. Bertani, \textit{Experimental realization of any discrete unitary operator}, Phys. Rev. Lett. \textbf{73}, 58-61 (1994).

\bibitem{R22} M. Zukowski, A. Zeilinger, and M. Horne, \textit{Realizable higher-dimensional two-particle entanglements via multiport beam splitters}, Phys. Rev. A \textbf{55}, 2564-2579 (1997).

\bibitem{R23} L. Allen, M. W. Beijersbergen, R. J. C. Spreeuw, and J. P. Woerdman, \textit{Orbital angular momentum of light and the transformation of Laguerre-Gaussian laser modes}, Phys. Rev. A \textbf{45}, 8185 (1992).

\bibitem{R24} J. Leach, B. Jack, J. Romero, A. K. Jha, A. M. Yao, S. Franke-Arnold, D. G. Ireland, R. W. Boyd, S. M. Barnett, and M. J. Padgett, \textit{Quantum correlations in optical angle-orbital angular momentum variables}, Science \textbf{329}, 662 (2010); S. Franke-Arnold, S. M Barnett, E. Yao, J. Leach, J. Courtial, and M. Padgett, \textit{Uncertainty principle for angular position and angular momentum}, New J. Phys. \textbf{6}, 103 (2004).

\bibitem{R25} A. Vaziri, G. Weihs, and A. Zeilinger, \textit{Experimental two-photon, three-dimensional entanglement for quantum communication}, Phys. Rev. Lett. \textbf{89}, 240401 (2002); J. Leach , B. Jack, J. Romero, M. Ritsch-Marte, R. W. Boyd, A. K. Jha, S. M. Barnett, S. Franke-Arnold, and M. J. Padgett, \textit{Violation of a Bell inequality in two-dimensional orbital angular momentum state-spaces}, Opt. Express \textbf{17}, 8287 (2009).

\bibitem{R26} M. Agnew, J. Z. Salvail, J. Leach, and R. W. Boyd, \textit{Generation of orbital angular momentum Bell states and their verification via accessible nonlinear witnesses}, Phys. Rev. Lett. \textbf{111}, 030402 (2013).

\bibitem{R27} N. K. Langford, R. B. Dalton, M. D. Harvey, J. L. O'Brien, G. J. Pryde, A. Gilchrist, S. D. Bartlett, and A. G. White, \textit{Measuring entangled qutrits and their use for quantum bit commitment}, Phys. Rev. Lett. \textbf{93}, 053601 (2004).

\bibitem{R28} J. B. Pors, S. S. R. Oemrawsingh, A. Aiello, M. P. van Exter, E. R. Eliel, G. W. 't Hooft, and J. P. Woerdman, \textit{Shannon dimensionality of quantum channels and its application to photon entanglement}, Phys. Rev. Lett. \textbf{101}, 120502 (2008).

\bibitem{R29} A. Mair, A. Vaziri, G. Weihs, and A. Zeilinger, \textit{Entanglement of the orbital angular momentum states of photons}, Nature \textbf{412}, 313 (2001).

\bibitem{R30} A. C. Dada, J. Leach, G. S. Buller, M. J. Padgett, and E. Andersson, \textit{Experimental high-dimensional two-photon entanglement and violations of generalized Bell inequalities}, Nature Phys. \textbf{7}, 677 (2011); B. J. Pors, F. Miatto, G. W. 't Hooft, E. R. Eliel, and J. P. Woerdman, \textit{High-dimensional entanglement with orbital-angular-momentum states of light}, J. Opt. \textbf{13}, 064008 (2011).

\bibitem{R31} R. Fickler, R. Lapkiewicz, W. N. Plick, M. Krenn, C. Schaeff, S. Ramelow, and A. Zeilinger, \textit{Quantum entanglement of high angular momenta}, Science \textbf{338}, 640 (2012).

\bibitem{R32} J. T. Barreiro, T. C. Wei, and P. G. Kwiat, \textit{Remote preparation of single-photon ``hybrid" entangled and vector-polarization states}, Phys. Rev. Lett. \textbf{105}, 030407 (2010).

\bibitem{R33} A. Holleczek, A. Aiello, C. Gabriel, C. Marquardt, and G. Leuchs, \textit{Classical and quantum properties of cylindrically polarized states of light}, Opt. Express \textbf{19}, 9714 (2011).

\bibitem{R34} X. L. Wang, J. Ding, W. J. Ni, C. S. Guo, and H. T. Wang, \textit{Generation of arbitrary vector beams with a spatial light modulator and a common path interferometric arrangement}, Opt. Lett. \textbf{32}, 3549 (2007); X. L. Wang, Y. N. Li, J. Chen, C. S. Guo, J. P. Ding, and H. T. Wang, \textit{A new type of vector fields with hybrid states of polarization}, Opt. Express. \textbf{18}, 10786 (2010).

\bibitem{R35} Q. Zhan, \textit{Cylindrical vector beams: from mathematical concepts to applications}, Adv. Opt. Photon. \textbf{1}, 1 (2009).

\bibitem{R36} L. Marrucci, C. Manzo, and D. Paparo, \textit{Optical spin-to-orbital angular momentum conversion in inhomogeneous anisotropic media}, Phys. Rev. Lett. \textbf{96}, 163905 (2006); C. Maurer, A. Jesacher, S. Furhapter, S. Bernet, and M. Ritsch-Marte, \textit{Tailoring of arbitrary optical vector beams}, New J. Phys. \textbf{9}, 78 (2007).

\bibitem{R37} S. M. Li, S. X. Qian, L. J. Kong, Z. C. Ren, Y. N. Li, C. H. Tu, and H. T. Wang, \textit{An efficient and robust scheme for controlling the states of polarization in a Sagnac interferometric configuration}, Europhys. Lett. \textbf{105}, 64006 (2014).

\bibitem{R38} S. Franke-Arnold, L. Allen, and M. Padget, \textit{Advances in optical angular momentum}, Laser Photon. Rev. \textbf{2}, 299 (2008).

\bibitem{R39} X. L. Wang, J. Chen, Y. N. Li, J. P. Ding, C. S. Guo, and H. T. Wang, \textit{Optical orbital angular momentum from the curl of polarization}, Phys. Rev. Lett. \textbf{105}, 253602 (2010); S. M. Li, Y. N. Li, X. L. Wang, L. J. Kong, K. Lou, C. H. Tu, Y. J. Tian, and H. T. Wang, \textit{Taming the collapse of optical fields}, Sci. Rep. \textbf{2}, 1007 (2012); K. Lou, S. X. Qian, Z. C. Ren, C. H. Tu, Y. N. Li, and H. T. Wang, \textit{Femtosecond laser processing by using patterned vector optical fields}, Sci. Rep. \textbf{3}, 2281 (2013).

\bibitem{R40} Q. Zhan, \textit{Trapping metallic Rayleigh particles with radial polarization}, Opt. Express \textbf{12}, 3377 (2004); R. Dorn, S. Quabis, and G. Leuchs, \textit{Sharper focus for a radially polarized light beam}, Phys. Rev. Lett. \textbf{91}, 233901 (2003).

\bibitem{R41} M. Sondermann, R. Maiwald, H. Konermann, N. Lindlein, U. Peschel, and G. Leuchs, \textit{Design of a mode converter for efficient light-atom coupling in free space}, Appl. Phys. B \textbf{89}, 489 (2007); H. P. Urbach and S. F. Pereira, \textit{Field in focus with a maximum longitudinal electric component}, Phys. Rev. Lett. \textbf{100}, 123904 (2008).

\bibitem{R42} B. R. Gadway, E. J. Galvez, and F. De Zela, \textit{Bell-inequality violations with single photons entangled in momentum and polarization}, J. Phys. B: At. Mol. Opt. Phys. \textbf{42}, 015503 (2009).

\end{references}
\end{document}